\documentstyle[11pt,psfig,aaspp4]{article}
\begin{document}

\title {THE MILLENNIUM ARECIBO 21-CM ABSORPTION LINE SURVEY: \\
EXTOLLING THE ANALYSIS PROCEDURES}

\author{ Carl Heiles}
\affil {Astronomy Department, University of California,
    Berkeley, CA 94720-3411; cheiles@astron.berkeley.edu}

\begin{abstract}

 	We have recently completed the analysis of a survey of the HI line in
emission/absorption against 79 radio sources.  Our CNM temperatures are
about half those of previous authors and we find a substantial quantity
of thermally unstable WNM.  These results are somewhat controversial. 
Here we discuss the analysis procedures to illustrate explicitly why the
results are reliable.

\end{abstract}

	Our full papers (Heiles \& Troland 2002a, 2002b) discuss our
analysis procedures and results in complete detail.  The main
observational results are \begin{itemize}

	\item The Cold Neutral Medium and Warm Neutral Medium (CNM and
WNM) are physically distinct components, existing in the ratio $\sim
40:60$. 

	\item Previously-derived CNM spin temperatures are typically too
high by a factor of two; our CNM temperature histogram peaks at about 40
K. The difference arises mainly from our including a proper radiative
transfer treatment.

	\item About $50\%$ of the WNM has upper limits to the kinetic
temperature $T_{kmax}$ in the thermally unstable region 500 to 5000 K. 
These are derived from line widths, which are in turn derived from
Gaussian components.  \end{itemize}

	The last item is particularly important because it violates the
cornerstone of most ISM models, such as McKee and Ostriker (1977), that
the WNM's temperature is in steady-state equilibrium between microscopic
heating and cooling processes.  However, our $T_{kmax}$ values come from
Gaussian-fit line widths.  Because of the difficulties with Gaussian
fitting our results have been criticized as being arbitrary, capricious,
subjective, nonunique, and of doing ``the field of ISM physics great
damage''.  Consequently, it seems worth spending our limited space
illustrating our fitting process to show that there is, in fact, little
room for doubt in the WNM Gaussians. 

\section{OBTAINING THE OPACITY AND EXPECTED HI PROFILES}

	One can derive spin temperature $T_s$ from emission-absorption
observations because the opacity $\tau(\nu)$ depends on $T_s$
differently than the brightness temperature $T_B(\nu)$ does.  Naively, one
makes one ON-source and one or more OFF-source observations.  One
combines the off-source observations to produce the expected profile
$T_{exp}(\nu)$, which is the $T_B(\nu)$ that would be seen at the source
position if the source were magically turned off.  Both $\tau(\nu)$ and
$T_{exp}(\nu)$ are affected by HI structure on the sky; these
uncertainties are a major item. 

	For each source we observed a grid of 16 OFF-source positions,
plus the ON-source position.  This allows us to expand the observed
brightness temperature $T_B({\nu})$ in a second-order two-dimensional
Taylor series around the source position and to derive the following
quantities, {\it plus their uncertainties}: \begin{enumerate}

	\item The opacity profile $\tau(\nu)$.

	\item The expected profile $T_{exp}(\nu)$.

	\item The first-derivative profiles ${\partial T_B(\nu) \over
\partial \alpha}$ and ${\partial T_B(\nu) \over \partial \delta}$
($\alpha$ is right ascension and $\delta$ is declination). 

	\item The second-derivative profiles ${\partial^2 T_B(\nu) \over
\partial \alpha^2}$, ${\partial^2 T_B(\nu) \over \partial \alpha
\partial \delta}$, and ${\partial^2 T_B(\nu) \over \partial \delta^2}$. 

\end{enumerate}

\begin{figure}[h!]
\begin{center}
\leavevmode
\epsfxsize=3in
\epsffile {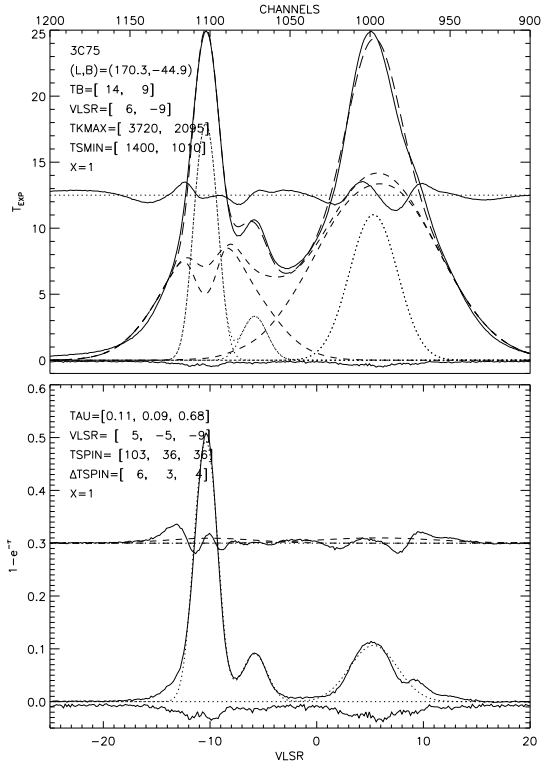}
\end{center}

\caption{Two-panel plot for 3C75.  For the complete description, see the
text.  \label{3C75}} \end{figure}

	Figure \ref{3C75} exhibits results for 3C75.  In the top panel,
the solid line profile extending above zero is $T_{exp}(\nu)$.  The
dashed curves show the observed emission from each individual WNM
Gaussian, including absorption by the CNM, and the heavy dashed curve
shows the totality of WNM fitted Gaussians.  The dotted curves show the
intrinsic emission from each individual CNM component, and the heavy
dots the observed emission of all CNM components including absorption
from the CNM components that lie in front.  The solid line near
mid-profile-height is the residuals of the data from the fit.  The solid
line profile extending below zero is the (negative of the) uncertainties
in $T_{exp}(\nu)$.  The annotation shows the properties of the WNM
Gaussian components.

	In the bottom panel, the solid line profile extending above zero
is the opacity profile $[1 - e^{\tau(\nu)}]$.  The dotted curves show
each individual CNM opacity Gaussian and the heavy dots show the opacity
sum of all the CNM Gaussians ($[1 - e^{- \sum\tau (\nu)}]$.  The solid
line near mid-profile-height is the residuals of the data from the fit;
the dashed line shows the contribution to opacity from each WNM
component assuming that its temperature is $T_{kmax}$.  The solid line
profile extending below zero is the (negative of the) uncertainty in the
opacity profile.  The annotation shows the properties of the CNM
Gaussian components. 

\section{THE GAUSSIAN ANALYSIS PROCEDURE}

	We choose 3C75 as an example because both the opacity and
expected profiles are fairly complicated, yet the fits are unambiguous. 
There are many sources with less complicated profiles that are fit with
fewer components.  Of the 79 sources, 58 have two or WNM fewer
components.  

	In the fitting process we first fit the opacity profile with the
fewest number of CNM Gaussian components required to make the residuals,
on the middle line, comparable with the uncertainties, the
negative-going profile at the bottom.  For 3C75 this requires three CNM
Gaussians, one for each of the major peaks.  They are unambiguous: one
could certainly not fit fewer.  One could include one or more additional
components, but this is not required in view of the uncertainties.  In
the main papers we show that this arbitrariness does not affect derived
spin temperatures. 

	We next perform a least squares fit on $T_{exp}(\nu)$.  The
parameters include one spin temperature for each CNM component; emission
from an arbitrary number of WNM components; the ordering of the CNM
components along the line of sight; and the fraction of WNM that lies
behind the CNM components.  These last two are required because the CNM
components absorb emission from WNM that lies behind them, and each CNM
component absorbs emission from the CNM components that lie behind it. 

	It is important to note that the CNM emission, shown by the
dotted lines on the top panel, is almost always narrower than the total
emission.  This highlights the {\it requirement} for WNM emission that
is both broader than the CNM and, also, produces no detectable
absorption.  This width difference means that: \begin{itemize}

	\item The spin temperatures depend mainly on the $T_{exp}(\nu)$
intensity {\it within} the velocity ranges covered by the dotted-line
CNM opacity Gaussians. 

	\item The properties of the WNM Gaussians are determined mainly
by the $T_{exp}(\nu)$ profile shape lying {\it outside} the velocity
ranges covered by the dotted-line CNM opacity Gaussians.  \end{itemize}

	In this case of 3C75, we derive upper limits to WNM kinetic
temperature $T_{kmax} = 3720$ and 2095 K; both are in the thermally
unstable range.  Looking at the $T_{exp}(\nu)$ profile, it is clear that
there is no way to obtain {\it wider} WNM Gaussians; the WNM widths are
constrained by the portions of the emission profile that lie outside
the dotted CNM emission.  One can obtain {\it narrower} WNM Gaussians in
two ways: \begin{enumerate}

	\item Include a very broad, weak Gaussian to sit underneath the
main emission profile. This would eliminate the systematic residuals in
the profile wings and produce a better fit. We included such a component
for many sources; indeed, for most sources with two WNM components, one
of the components is usually such a broad, weak pedestal.

	\item Fit a larger number of narrow WNM Gaussians to
$T_{exp}(\nu)$.  One can often replace a single-Gaussian fit by more
narrower Gaussians.  However, this violates the philosophy of using the
minimum number of free parameters.  Moreover, with more narrower WNM
components, the WNM temperatures would decrease and their associated
opacities would increase; given the uncertainties in $\tau(\nu)$ there
is some room for this, but not much.  \end{enumerate}

	In this case, then, we conclude that WNM lying in the unstable
temperature range is undeniable. Similar conclusions apply in most
cases. Some sources have less unambiguous fitting results, but we always
attempt to be ``reasonable'' and this philosophy leads to the statistics
that we quote in the abstract and introduction.

\acknowledgements

	This work was supported in part by NSF grants AST-9530590 and
AST-0097417, and by the NAIC.

\end{document}